\documentclass[twocolumn,english]{revtex4-1}
\usepackage[LGR,T1]{fontenc}
\usepackage[latin9]{inputenc}
\usepackage{amsmath}
\usepackage{amssymb}
\usepackage{graphicx}
\usepackage{wasysym}

\makeatletter

\newcommand*\LyXThinSpace{\,\hspace{0pt}}
\DeclareRobustCommand{\greektext}{%
  \fontencoding{LGR}\selectfont\def\encodingdefault{LGR}}
\DeclareRobustCommand{\textgreek}[1]{\leavevmode{\greektext #1}}
\ProvideTextCommand{\~}{LGR}[1]{\char126#1}

\makeatother

\usepackage{babel}
\begin{document}
\title{Probing Primordial Symmetry Breaking with Cosmic Microwave Background
Anisotropy}
\author{Jiwon Park}
\email{cosmosapjw@soongsil.ac.kr}

\author{Tae Hoon Lee}
\affiliation{Department of Physics, Soongsil University, Seoul 06978, Korea}
\begin{abstract}
There have been vigorous research attempts to test various modified
gravity theories by using physics of the cosmic microwave background
(CMB). Meanwhile, symmetry breaking such as Higgs mechanism is one
of the most important phenomena in physics but there have been not
so much researches to make them contact with cosmological observations.
In this article, with the CMB power spectra we try to distinguish
two different scenarios of spontaneous symmetry breaking in primordial
era of the universe. The first model is based on a broken symmetric
theory of gravity, which was suggested by A. Zee in 1979. The second
model is an application of Palatini formalism to the first model.
Perturbation equations are computed and they show differences originated
from the property of symmetry. Furthermore, it turns out that two
models have different features of CMB power spectra with the same
potential scale. This fact enables us to verify distinct kinds of
primordial symmetry breaking with CMB physics.
\end{abstract}
\maketitle

\section{Introduction}

After the discovery of Cosmic Microwave Background (CMB), there have
been many studies to understand its physical implications. Especially,
its anisotropy has been vigorously studied to verify and give restriction
on free parameters of modified gravity theories such as Brans-Dicke
(BD) theory {[}1{]}, Horndeski theory {[}2{]}, and $f(R)$ gravity
{[}3{]}. They have been developed to alleviate or give a clue for
cosmological problems such as the cosmic constant problem.

Among the modified theories, there also have been attempts to apply
an idea of symmetry breaking to cosmology. The studies on symmetries
and their breaking phenomena is one of the most important development
in modern physics. The idea of spontaneous symmetry breaking, especially
after the appearance of Higgs field theory {[}4-6{]}, became one of
the main subject area in particle physics. This trend also has affected
studies in cosmology such as cosmic acceleration, the theory of inflation
and the cosmic constant problem {[}7{]}. There are other tries for
doing it in the primordial era. Broken-symmetric theory of gravity
proposed by Zee {[}8{]} is a remarkable one among these attempts,
which applies Higgs-type mechanism in BD theory with the potential.
In the theory, the potential invokes symmetry breaking in primordial
era so that the BD field value settle down at Planck mass. One of
the main features of the model is that it cannot be distinguished
from general relativity (GR) by current observation, if we consider
only non-perturbative scale and recent era of the universe.

This model gives us deep insights, however, its properties on perturbative
scale were not so much studied. One of the preceding researches studying
perturbations in Zee's theory given by {[}9{]} only discusses the
simple solution for perturbation. Moreover, it is focused on the view
of particle physics and deals no specific method to verify the theory
with cosmological tests. In the paper {[}10{]}, the perturbation of
the scalar field is mentioned but it is only used for showing some
problems which may arise in the open universe. In another study {[}11{]},
some effects in CMB which the field may bring are briefly mentioned.
However, these are not highlighted and then an alternative scenario
to use the field as inflaton is soon discussed. Therefore, we are
motivated to find how we can verify the theory with cosmological observations.
In this paper, we revisit broken-symmetric theory of gravity, by studying
how it affects CMB anisotropy spectra. Furthermore, we also study
its modification, to investigate whether a different type of symmetry
breaking may bring differences in observables.

Our modification is based on Palatini formalism. In GR, for the uniqueness
of a connection we assume metric compatibility, which says that a
covariant derivative of the metric tensor is zero. From this assumption
we can compute Affine connection from the metric tensor. In contrast,
in the Palatini formalism it is assumed that the connection is independent
from the metric tensor. Especially, when we apply it to the modified
gravity theories, metric compatibility may also not be true anymore.
There are two main application of Palatini formalism in modified gravity
theories. They are known as Weyl scalar-tensor geometry (or Weyl geometry)
{[}12{]} and Palatini $f(R)$ gravity {[}13{]}, which are their adoptions
to BD theory and $f(R)$ gravity. Here we propose a slightly different
model for broken-symmetric theory of gravity by using Weyl geometry,
and investigate whether one can distinguish two different types of
primordial symmetry breaking by perturbation theory.

Throughout the paper, to construct perturbation theory we use covariant
and gauge-invariant formalism (or covariant formalism), which was
developed by {[}14-16{]}. The reason why we choose covariant formalism
is that we choose CAMB {[}17{]} to execute numerical analysis. CAMB
takes covariant formalism to be its fundamental numerical stretegy,
due to the well-known fact that there is a gauge choosing problem
in cosmological perturbation theory (see {[}18{]} for detailed explanation,
for example). Covariant formalism gives a simple remedy by using 1+3
decomposition of Einstein field equation and using a fundamental 4-velocity
of fluid as a source for basic kinematical quantities which are always
gauge-independent. It has a clear physical meaning and its equations
are equivalent to synchronous gauge if we set the observer 4-velocity
to correspond to cold dark matter (CDM) velocity.

This paper is organized as follows: In section II we review broken-symmetric
theory of gravity and Weyl geometry, then we construct our new kind
of broken-symmetric theory of gravity by using Palatini formalism.
We show that one cannot verify these models only by studying non-perturbative
scale. In section III we construct linearized perturbation theory
based on covariant formalism, and derive additional energy-momentum
(EM) tensor in perturbative scale for each model. Then we discuss
differences between two models in perturbative scale. In section IV,
to perform numerical analysis, first we propose approximation scheme
for the scalar field, for the numerical code CAMB. Next, we explain
some new features in the perturbation of energy density for ordinary
matter. This helps us when we try to understand the properties of
CMB power spectra. We analyze numerical results for CMB power spectra.
Then we compare our final results with other researches, and discuss
some observational tests that would be helpful when verifying our
models. In section V we summarize our results and discuss future prospects
and limits of our study.

\section{two models of primordial symmetry breaking in cosmology}

We first begin this section by reviewing Zee's broken-symmetric theory
of gravity, and we propose our new model by applying Palatini formalism
to Zee's theory. Basically, two models both adopt the action of BD
theory with potential for their actions. The potential breakes down
the symmetry so that the BD scalar field is only detectable in the
perturbative scale. However, two models are expected to bring different
effects since these have different symmetry. We investigate this effects
for each model. For a briefness, we call each models as the model
A and the model B from now on.

The action of BD theory with potential is given by

\begin{equation}
S=\int\mathrm{d}^{4}x\sqrt{-g}\,[\frac{1}{2}\phi^{2}R+\frac{1}{2}g^{\alpha\beta}\nabla_{\alpha}\phi\nabla_{\beta}\phi+V(\phi)]+S_{\mathrm{M}},
\end{equation}
where $R$ is Ricci scalar (for the definition of Ricci scalar and
related quantities, see (25) and comment below), $S_{\mathrm{M}}$
is the action for ordinary matter. In the model A, the following Higgs-type
potential is adopted to invoke symmetry breaking:
\begin{equation}
V(\phi)\equiv\frac{1}{2}V_{\mathrm{A}}(\phi^{2}-M_{\mathrm{P}}^{2})^{2},
\end{equation}
where $M_{\mathrm{P}}$ is Planck mass and $V_{\mathrm{A}}$ is a
scale for the potential. Let us adopt Levi-Civita connection, which
is given by metric compatibility condition $\nabla_{\gamma}g_{\mu\nu}=0$,
to be connection in the model A. Then the equations of motion obtained
from the action (1) are given by
\begin{equation}
G_{\mu\nu}=\frac{2}{\phi^{2}}T_{\mu\nu}+T_{\mu\nu}^{(\phi)},
\end{equation}
\begin{equation}
\square\phi-\partial_{\phi}V(\phi)-\phi R=0,
\end{equation}
where $G_{\mu\nu}\equiv R_{\mu\nu}-g_{\mu\nu}R/2$, $\square\equiv\nabla^{\mu}\nabla_{\mu}$,
$R_{\mu\nu}$ is Ricci tensor and we define EM tensor for matter and
$\phi$:
\begin{equation}
T_{\mu\nu}\equiv-\frac{2}{\sqrt{-g}}\frac{\delta S_{\mathrm{M}}}{\delta g^{\mu\nu}},
\end{equation}
\begin{align}
T_{\mu\nu}^{(\phi)}\equiv & -\frac{2}{\phi^{2}}\partial_{\mu}\phi\partial_{\nu}\phi+\frac{2}{\phi^{2}}g_{\mu\nu}[\frac{1}{2}\partial_{\alpha}\phi\partial^{\alpha}\phi+V(\phi)]\nonumber \\
 & +\frac{1}{\phi^{2}}(\nabla_{\mu}\nabla_{\nu}\phi^{2}-g_{\mu\nu}\square\phi^{2}).
\end{align}
From equation (4), we see that the potential given by (2) invokes
symmetry breaking and make the field $\phi$ settle down at the minimum
$\phi=M_{\mathrm{P}}$. After the symmetry breaking, the theory recovers
GR in non-perturbative scale so that
\begin{equation}
G_{\mu\nu}=\kappa T_{\mu\nu},
\end{equation}
where $\kappa=2/M_{\mathrm{P}}^{2}$. 

The model B shares almost the same mechanism but it differs from the
model A in detail. In the model A we set the connection to be Levi-Civita
connection, which is exactly the same connection in GR. We can derive
it from the metric tensor. In the model B, however, the connection
is considered to be independent from the metric. This condition is
known as Palatini formalism. From this new formalism we can derive
a equation on this new connection, namely $\bar{\Gamma}_{\mu\nu}^{\alpha}$.
Taking a variation of the action (1) with respect to $\bar{\Gamma}_{\mu\nu}^{\alpha}$,
we obtain
\begin{equation}
\bar{\nabla}_{\alpha}g_{\mu\nu}=g_{\mu\nu}\psi_{,\alpha},
\end{equation}
where $\bar{\nabla}_{\alpha}$ is a new covariant derivative corresponding
to $\bar{\Gamma}_{\mu\nu}^{\alpha}$ and we define a rescaling of
the field $\psi=-2\ln\phi$ for convenience. Now we set the derivative
operator $\bar{\nabla}_{\alpha}$ satisfying (8) to be fundamental
derivative operator in the model B, instead of original covariant
derivative $\nabla_{\alpha}$ which satisfies metric compatibillity.
We call this kinds of theories with the covariant derivative given
by (8) as Weyl geometry, whose name is originated from Herman Weyl
{[}19{]}. Likewise, we call the new connection $\bar{\Gamma}_{\mu\nu}^{\alpha}$
satisfying (8) Weyl connection. It is related with the original Levi-Civita
connection $\Gamma_{\mu\nu}^{\alpha}$ as
\begin{equation}
\bar{\Gamma}_{\mu\nu}^{\alpha}=\Gamma_{\mu\nu}^{\alpha}+\delta\Gamma_{\mu\nu}^{\alpha},
\end{equation}
where
\begin{equation}
\delta\Gamma_{\mu\nu}^{\alpha}\equiv-\frac{1}{2}(g_{\phantom{\alpha}\mu}^{\alpha}\psi_{,\nu}+g_{\phantom{\alpha}\nu}^{\alpha}\psi_{,\mu}-g_{\mu\nu}\psi^{,\alpha}).
\end{equation}

Before we do more explicit computation, let us explain its geometric
characters for the later conveniences. The spacetime with Weyl connection
has new kind of symmetry, because (8) is invariant under the transformation
\begin{equation}
(g_{\mu\nu},\psi)\rightarrow(e^{f}g_{\mu\nu},\psi+f),
\end{equation}
where $f$ can be any function on spacetime. This kinds of transformation
consist a group, namely $G_{\mathrm{weyl}}(\psi)$. Any operation
in $G_{\mathrm{weyl}}$ is called a frame transformation and pair
$(M,g_{\mu\nu},\psi)$ is called a frame where $M$ is spacetime manifold.
Especially, if we take $f=-\psi$ and define $\gamma_{\mu\nu}\equiv e^{-\psi}g_{\mu\nu}$,
namely effective metric, then from (8) it is clear that $\bar{\nabla}_{\alpha}\gamma_{\mu\nu}=0$.
So for the covariant derivative operator in the model B, the metric
$\gamma_{\mu\nu}$ behave like original metric $g_{\mu\nu}$ for the
derivative operator in GR. Hence, we can take similar step like GR
for computation involving derivation if we use the rescaled metric
$\gamma_{\mu\nu}$ in the frame. We call the frame $(M,\gamma_{\mu\nu},0)$
a Riemann frame, whereas the frame with $g_{\mu\nu}$ is called a
Weyl frame.

It is convenient to rewrite the action in the Riemannian frame, for
the fact that in this frame the field $\psi$ appear to be minimally
coupled. We propose the action for the model B to be
\begin{equation}
S=\frac{M_{\mathrm{P}}^{2}}{2}\int\mathrm{d}^{4}x\sqrt{-\gamma}\,[\bar{R}+\frac{1}{8}\gamma^{\alpha\beta}\bar{\nabla}_{\alpha}\psi\bar{\nabla}_{\beta}\psi+e^{2\psi}V(\psi)]+\bar{S}_{\mathrm{M}},
\end{equation}
where
\begin{equation}
V(\psi)\equiv\frac{1}{2}V_{\mathrm{B}}M_{\mathrm{P}}^{2}(e^{-\psi}-v^{2})^{2}.
\end{equation}
Here the Planck constant appears explicitly because the field $\psi$
is now dimensionless and cannot play a role for the mass and a factor
$1/8$ in the kinetic term for $\psi$ comes from the rescaling of
the field. Ricci scalar $\bar{R}$ is also redefined with respect
to Weyl connection. The matter action $\bar{S}_{\mathrm{M}}$ is of
course changed to satisfy the symmetry condition (11) too. Note that
the Planck constant $M_{\mathrm{P}}$ in the potential is changed
to $v$, which has same value of $M_{\mathrm{P}}$ but is dimensionless
and the scale for the potential is also rescaled. From now on we denote
terms (or operators) with bar as redefined quantities with the new
connection in the model B.

The equations of motion from the action (12), in Riemannian frame,
are given by
\begin{equation}
\bar{G}_{\mu\nu}=\kappa\bar{T}_{\mu\nu}+\bar{T}_{\mu\nu}^{(\psi)},
\end{equation}
\begin{equation}
\bar{\square}\psi-\partial_{\psi}V_{\mathrm{eff}}(\psi)=0,
\end{equation}
where we redefine the EM tensor for matter and $\psi$:
\begin{equation}
\bar{T}_{\mu\nu}\equiv-\frac{2}{\sqrt{-\gamma}}\frac{\delta\bar{S}_{\mathrm{M}}}{\delta\gamma^{\mu\nu}},
\end{equation}
\begin{equation}
\bar{T}_{\mu\nu}^{(\psi)}\equiv-\partial_{\mu}\psi\partial_{\nu}\psi+\frac{1}{2}\gamma_{\mu\nu}[\gamma^{\alpha\beta}\partial_{\alpha}\psi\partial_{\beta}\psi+e^{2\psi}V(\phi)],
\end{equation}
and the effective potential
\begin{equation}
\partial_{\psi}V_{\mathrm{eff}}(\psi)\equiv4e^{2\psi}[\partial_{\psi}V(\psi)+2V(\psi)].
\end{equation}
Since $V_{\mathrm{eff}}^{\prime}(e^{-\psi}=v^{2})=0$ and $V_{\mathrm{eff}}^{\prime\prime}(e^{-\psi}=v^{2})>0$,
The potential also can invoke symmetry breaking at $e^{-\psi}=v^{2}$.

Before we discuss effects of the symmetry breaking in CMB anisotropy,
we explain a geometirc role of the symmetry breaking for deeper understanding.
As we have shown before, the action (12) is invariant under the operations
of the transformation group $G_{\mathrm{weyl}}(\psi)$. The invariant
quantity corresponds to this group is of course Weyl connection $\bar{\Gamma}_{\mu\nu}^{\alpha}$.
After the symmetry breaking, on non-perturbative scale, the field
$\psi$ is fixed to be $\psi=-\ln v^{2}$ so spacetime symmetry is
now broken. On the perturbative scale, however, there arises a new
symmetry whose operation is given by
\begin{equation}
(g_{\mu\nu},\zeta)\rightarrow(e^{f}g_{\mu\nu},\zeta+f),
\end{equation}
where we expand $\psi=-\ln v^{2}+\zeta$ with the condition $\left|\zeta\right|\ll\left|\ln v^{2}\right|$.
Now the spacetime symmetry obeys new group $G_{\mathrm{weyl}}(\zeta)$
on the perturbative scale and thus we may call symmetry breaking in
our model geometrical. This illustrates why we should consider perturbation
theory in the model B after the symmetry breaking. Because of the
geometrical feature of the field $\psi$, all the geometrical observables
must be defined to be invariant under $G_{\mathrm{weyl}}(\psi)$.
After the symmetry breaking, however, on the non-perturbative scale
there is no geometrical effects of $\psi$ anymore. That is to say,
the group $G_{\mathrm{weyl}}(\psi)$ has only one elements, the identity.
Therefore after the symmetry breaking we can see the effects of $\psi$
only through the perturbation scale, where the group $G_{\mathrm{weyl}}(\zeta)$
has nontrivial elements. In the next section we investigate these
perturbative features in more detail with covariant formalism.

\section{covariant formalism and perturbation theory}

Here we illustrate covariant formalism which is used by the numerical
code CAMB to explain perturbative features of two models. The formalism
uses 1+3 decomposition of the spacetime. For given coordinates $x_{a}$,
we define a 4-velocity
\begin{equation}
u_{a}\equiv\frac{\mathrm{d}x_{a}}{\mathrm{d}\tau},
\end{equation}
with a comoving time $\tau$ {[}20{]}. We choose 4-velocity $u_{a}$
to be $u_{a}u^{a}=1$. With $u_{a}$ one can make 1+3 decomposition.
We split the covariant derivative $\nabla$ to make time derivative
and orthogonal spatial derivative as
\begin{equation}
\dot{S}_{\phantom{a\dots}b\dots}^{a\ldots}\equiv u^{c}\nabla_{c}S_{\phantom{a\dots}b\dots}^{a\ldots},
\end{equation}
\begin{equation}
D_{c}S_{\phantom{a\dots}b\dots}^{a\ldots}\equiv h_{c}^{\phantom{c}f}h_{\phantom{a}d}^{a}\cdots h_{b}^{\phantom{b}e}\cdots\nabla_{f}S_{\phantom{d\cdots}e\cdots}^{d\cdots},
\end{equation}
where $h_{ab}\equiv g_{ab}-u_{a}u_{b}$ is spatial projection tensor.
Note that here we only give definitions with the Levi-Civita connection.
The reason why we do not need one with Weyl connection wiil be discussed
soon. Another useful definition is projected symmetric trace-free
(PSTF) part of vectors and tensors. The PSTF parts of vector and tensor
are given by
\begin{equation}
V_{\langle a\rangle}\equiv h_{ab}V^{b},
\end{equation}
\begin{equation}
S_{\langle ab\dots z\rangle}\equiv(h_{(a}^{\phantom{a}c}h_{b}^{\phantom{b}d}\cdots h_{z)}^{\phantom{b}w}-\frac{1}{3}h^{cd\dots w}h_{ab\dots z})S_{cd\dots w},
\end{equation}
where the bracket in the subscript denotes anti-commutator. With 4-velocity
we can also define Riemann tensor with the equation

\begin{equation}
(\nabla_{c}\nabla_{d}-\nabla_{d}\nabla_{c})u_{a}=R_{abcd}u^{b},
\end{equation}
from which we define Ricci tensor $R_{\mu\nu}\equiv R_{\gamma\mu\phantom{\gamma}\nu}^{\phantom{\gamma\mu}\gamma\phantom{\nu}}$
and Ricci scalar $R\equiv R_{\mu}^{\phantom{\mu}\mu}$. Next we split
EM tensor by using 1+3 decomposition:
\begin{equation}
T_{ab}=\rho u_{a}u_{b}+2q_{(a}u_{b)}-ph_{ab}+\pi_{ab},
\end{equation}
where $\rho\equiv T_{ab}u^{a}u^{b}$ is energy density, $q_{a}\equiv T_{\left\langle a\right\rangle b}u^{b}$
is momentum density, $p\equiv-h^{ab}T_{ab}/3$ is pressure, $\pi_{ab}\equiv T_{\left\langle ab\right\rangle }$
is anisotropic stress. At the zero-th order (non-perturbative scale),
for the isotropic and homogeneous universe all the quantities become
zero except for the energy density and the pressure.

Now we are ready to present perturbation equations for each model.
First we discuss the model A. Whereas these two models have exactly
same background equations as GR, on the perturbative scale differences
arise. Clearly, one can see that the additional EM tensor $T_{\mu\nu}^{(\phi)}$
for the scalar field $\phi$ does not totally vanish in first order.
The ordinary matter EM tensor $2T_{\mu\nu}/\phi^{2}$ also gives an
additional term. Therefore on the perturbative scale (3) become
\begin{equation}
\delta G_{\mu\nu}=\kappa\delta T_{\mu\nu}+\delta T_{\mu\nu}^{(\mathrm{A})}.
\end{equation}
Here the quantity with $\delta$ denotes its first order perturbation
and the additional EM tensor $\delta T_{\mu\nu}^{(\mathrm{A})}$,
which originated from the BD scalar field, is given by 
\begin{equation}
\delta T_{\mu\nu}^{(\mathrm{A})}\equiv\nabla_{\mu}\nabla_{\nu}\varphi_{\mathrm{A}}-g_{\mu\nu}\square\varphi_{\mathrm{A}}-\kappa\varphi_{\mathrm{A}}T_{\mu\nu},
\end{equation}
where $\varphi_{\mathrm{A}}\equiv2\delta\phi/M_{\mathrm{P}}$. From
this new EM tensor one may compute terms such as energy density, that
is given by (26). The evolution equations for $\varphi_{\mathrm{A}}$
is given by
\begin{equation}
\ddot{\varphi}_{\mathrm{A}}+3H\dot{\varphi}_{\mathrm{A}}+a^{2}D^{2}\varphi_{\mathrm{A}}+4V_{\mathrm{A}}M_{\mathrm{P}}^{2}\varphi_{\mathrm{A}}=\kappa(\delta\rho-3\delta p),
\end{equation}
where $a$ is a scale factor of the universe, $H\equiv\dot{a}/a$,
and $D^{2}\equiv D_{\mu}D^{\mu}$.

For the perturbation theory in the model B, we need to be more careful.
For an illustration, let $X^{\mu}$ and $Y^{\mu}$ be vector fields.
Suppose that $X^{\mu}$ is a zero-th order background function which
only depends on time but may depend on space location on the perturbative
scale. Hence we may regard its spatial gradient $\bar{D}_{\mu}X^{\mu}$
as a perturbative quantity. On the other hand, suppose that $Y^{\mu}$
is itself a first order perturbative variable. For consistency and
convenience, many times it is useful to regard $\bar{D}_{\mu}Y^{\mu}$
as a perturbative quantity, not $Y^{\mu}$ itself. However, the derivative
operators in two quantities in fact cannot be same in first order
scale. To be explicit, we obtain
\begin{equation}
\bar{D}_{\mu}X^{\mu}=D_{\mu}X^{\mu}-h^{\mu\nu}\delta\Gamma_{\mu\nu}^{\gamma}X_{\gamma},
\end{equation}
\begin{equation}
\bar{D}_{\mu}Y^{\mu}=D_{\mu}Y^{\mu},
\end{equation}
where the additional factor $-h^{\mu\nu}\delta\Gamma_{\mu\nu}^{\gamma}Y_{\gamma}$
due to Weyl connection vanishes in (31), because it is second order
term. Hence we do not use $\bar{\nabla}$ but rather only $\nabla$.
For the same reason from now on we only use Weyl frame.

Transforming $\bar{G}_{\mu\nu}$ into $G_{\mu\nu}$ and Considering
perturbation up to first order, we find 
\begin{equation}
\delta G_{\mu\nu}=\kappa\delta\bar{T}_{\mu\nu}+\delta T_{\mu\nu}^{(\mathrm{B})},
\end{equation}
where
\begin{equation}
\delta T_{\mu\nu}^{(\mathrm{B})}\equiv\nabla_{\mu}\nabla_{\nu}\varphi_{\mathrm{B}}-g_{\mu\nu}\square\varphi_{\mathrm{B}},
\end{equation}
and $\varphi_{\mathrm{B}}\equiv\delta\psi$. The evolution equations
for $\varphi_{\mathrm{B}}$ is given by
\begin{equation}
\ddot{\varphi}_{\mathrm{B}}+3H\dot{\varphi}_{\mathrm{B}}+a^{2}D^{2}\varphi_{\mathrm{B}}+4V_{\mathrm{B}}M_{\mathrm{P}}^{2}\varphi_{\mathrm{B}}=0.
\end{equation}
In the context of the field theory, one may define a mass of $\varphi_{\mathrm{A}}$
(or $\varphi_{\mathrm{B}}$) as $m_{\mathrm{A}}^{2}\equiv4V_{\mathrm{A}}M_{\mathrm{P}}^{2}$
(or $m_{\mathrm{B}}^{2}\equiv4V_{\mathrm{B}}M_{\mathrm{P}}^{2}$).
This definitions would be helpful in the view of particle physics,
especially in the next section.

In summary, we explain some differences between our two models A,
B and other researches. In general, when we consider the perturbation
theory in many modified gravity theories one has to consider not only
perturbation equations but also background equations. In our models,
however, we only need to consider the modified perturbation equations,
that is the new EM tensor perturbation which is given by (28) and
(33). This is quite noticeable difference between many other researches
on CMB anisotropy in modified gravity theories and our models. Our
models can be thought to be equivalent to GR as their efective theory
because the scalar field is decoupled after the symmetry breaking.
Note that, however, this effective theory becomes standard $\Lambda$CDM
model with the correction of a new additional EM tensor, which arises
from non-minimal coupling in the original action. Here the field mass
$m$ plays a crucial role in our models, as it determines how much
the models deviate from GR. Moreover, the new features of perturbations
represented by (28) and (33) reveal a difference between two models
explicitly. In (28), the new EM tensor contains term $\kappa\varphi_{\mathrm{A}}T_{\mu\nu}$
which comes from the EM tensor of ordinary matter, whereas in (33)
there is no such term.

This comes from the properties of symmetry which the action satisfies.
In the model A, the matter action $S_{\mathrm{M}}$ has no need to
couple with the scalar field $\phi$ explicitly in Weyl frame, since
there is no conformal symmetry like (11) to be satisfied. In the model
B, however, the matter action need to satisfy a new kind of symmetry
(19) and to satisfy the condition it must appear to be not coupled
with the field in Riemannian frame. This causes different features
of CMB power spectra in the model B from the model A, which we explain
from now on. To do this, let us discuss our final results of two models
with some numerical analysis.

\section{numerical analysis and results}

In this section, we present our numerical results obtained by a modification
of CAMB. CAMB is written in fortran 90, fast and accurate enough for
the statistical analysis of many data, and hence it is generally used
to analyze new CMB physics. To obtain numerical results, we modified
the set of equations in \texttt{equations.f90} file in folder \texttt{fortran}. 

Although CAMB can solve equations enough exactly, as the potential
scale become higher the field oscillates more rapidly in its restrictive
region (the field should be diluted fast in the radiation epoch, so
that the field cannot be detected in the current era), and hence to
solve many other coupled equations exactly CAMB needs to split the
region with extremely small time spacing to reflect the rapid changes
of the field, which results the code to be much slower than the usaul
cases {[}21{]}. Unfortunately, we could not avoid this numerical issue
in both models and we had to develop an approximation scheme for the
rapidly oscillating field. Our scheme is not to be quantitively accurate
but to cut down the time-consuming computation of the code and to
analyze intrinsic features of the physics qualitively.

In many times it is convenient to expand variables as series with
harmoric coefficients. We define scalar eigenfunction $Q(k)$ which
satisfies generalized Helmholtz equation 
\begin{equation}
a^{2}D^{2}Q(k)=k^{2}Q(k),
\end{equation}
and $\dot{Q}(k)=0$. With the definition
\begin{equation}
Q_{a_{1}a_{2}\ldots a_{l}}(k)\equiv(\frac{a}{k})^{l}D_{a_{1}}D_{a_{2}}\dots D_{a_{l}}Q(k),
\end{equation}
we may expand some quantity $X_{a_{1}a_{2}\dots a_{l}}$ as 
\begin{equation}
X_{a_{1}a_{2}\dots a_{l}}=\sum_{k}C_{\mathrm{X}}(k,a)X_{k}Q_{a_{1}a_{2}\ldots a_{l}}(k),
\end{equation}
where $C_{\mathrm{X}}(k,a)$ is a coefficient of the series that we
can set it for convenience. With these definitions we expand our scalar
field with harmonic quantities. First note that we may neglect the
terms in the right-hand side of (29), since the field would decrease
rapidly in the radiation era, so during the era that the field value
is considerable we may set $\delta(\rho+3p)\approx0$. Hence we may
define $\varphi\equiv\varphi_{\mathrm{A}}=\varphi_{\mathrm{B}}$ and
\begin{equation}
aD_{a}\varphi\equiv\sum_{k}k\phi_{k}Q_{a}(k).
\end{equation}
Then (29) and (34) become
\begin{equation}
\phi_{k}^{\prime\prime}+2\mathcal{H}\phi_{k}^{\prime}+k^{2}\phi_{k}+4a^{2}VM_{\mathrm{P}}^{2}\phi_{k}=0,
\end{equation}
where the prime denotes derivative with the conformal time $\mathrm{d}\tau\equiv\mathrm{d}t/a$,
$\mathcal{H}\equiv a^{\prime}/a$, and we omitted the subscript in
the potential scale for convenience. Now we propose our approximation
scheme. The field asymptotically decrease with a factor $1/a$ as
time goes by, so we may introduce a new variable $\varphi_{k}\equiv a\phi_{k}$
to conceal the feature. We can see that the equation for $\varphi_{k}$
shows the oscillating property more explicitly, with an analogy with
a simple harmonic oscillator:
\begin{equation}
\varphi_{k}^{\prime\prime}=-(k^{2}+4VM_{\mathrm{P}}^{2}a^{2}-\frac{a^{\prime\prime}}{a})\varphi_{k}.
\end{equation}
 We want to drop out the oscillation part and extract only effects
from the potential scale on the asymptotic line of the solution of
(39). Hence we decompose $\varphi_{k}=e^{g+ih}$, as a composition
of two real-valued variables $g$ which corresponds to amplitude and
$h$ which corresponds to oscillation and $i\equiv\sqrt{-1}$. The
equations for $g$ and $h$ are
\begin{equation}
g^{\prime\prime}=-(k^{2}+4VM_{\mathrm{P}}^{2}a^{2}-\frac{a^{\prime\prime}}{a})-(g^{\prime})^{2}+(h^{\prime})^{2},
\end{equation}
\begin{equation}
h^{\prime\prime}=-2g^{\prime}h^{\prime}.
\end{equation}
To rule out the effect of the oscillation, we assume that a wavelength
of $h$ is infinite. This implies $\left|h^{\prime}\right|\ll1$ and
$\left|g^{\prime}\right|\propto\left|h^{\prime\prime}/h^{\prime}\right|\ll1$.
Therefore, we can write the solution of $\phi_{k}$ as $\phi_{k}(\tau)\approx f(\tau)/a$,
where
\begin{equation}
f(\tau)\equiv\exp(-\int_{\tau_{0}}^{\tau}d\tau^{\prime\prime}\,\int_{\tau_{0}}^{\tau^{\prime\prime}}d\tau^{\prime}\,\{k^{2}+4VM_{\mathrm{P}}^{2}a^{2}-\frac{a^{\prime\prime}}{a}\}),
\end{equation}
where $\tau_{0}$ is an initial time when the code begins evolution.

Our approximation may seem to be somewhat handwavy, however, the most
important property of the evolution, its rapid decrease and dependency
on the potential scale, still survives. The oscillating feature mostly
dominates on relatively late-time era and its amplitude is quite small
and negligible so we expect that there is almost no danger to lose
physically important phenomena.

Next we discuss initial conditions for the scalar field. The simplest
choice for the initial values is of course $\phi_{k}=0$ and $\phi_{k}^{\prime}=0$,
but this values give no evolution to the field so do not have any
kind of interest. Rather we try to give a more physical condition.
For the early era of the universe and a long wavelenth region where
the code begins evolution, $\left|k\tau\right|\ll1$, (40) is reduced
to $\varphi_{k}^{\prime\prime}=(a^{\prime\prime}/a)\varphi_{k}$,
whose solution satisfies that $\varphi_{k}/a=\phi_{k}$ is constant.
Comparing coefficients in the actions of each model, one find $\phi_{k}=\sqrt{2}$
and $\phi_{k}^{\prime}=0$.

We are now ready to execute our numerical computation. For cosmological
parameters we use Planck 2018 results {[}22{]}: current Hubble factor
$H_{0}=67.32117\,\mathrm{Km\,\mathrm{s}^{-1}}\,\mathrm{Mpc}^{-1}$,
baryon density $\Omega_{\mathrm{b}}h^{2}=0.0223828$, cold dark matter
density $\Omega_{\mathrm{c}}h^{2}=0.1201075$, neutrino density $\Omega_{\mathrm{\nu}}h^{2}=0.6451439\times10^{-3}$,
scalar power spectra amplitude $A_{\mathrm{s}}=2.100549\times10^{-9}$,
and scalar spectral index $n_{\mathrm{s}}=0.9660499$, and we only
consider the flat universe.

Before we discuss CMB multipoles, let us explain effects of the additional
EM tensor on the evolution of matter perturbation. This will help
us to understand the features of our models more clearly. Consider
a barotropic matter with a constant equation of state $w\equiv p/\rho$.
The evolution equation of its energy density perturbation $\Delta_{k}$,
defined by $aD_{a}\rho=\rho\sum k\Delta_{k}Q_{a}(k)$, is given by
\begin{equation}
\Delta_{k}^{\prime\prime}+(1-3w)\mathcal{H}\Delta_{k}^{\prime}-[\frac{\kappa a^{2}}{2}(1+2w-3w^{2})\rho+k^{2}]\Delta_{k}=-F_{\mathrm{X}}(\phi_{k}),
\end{equation}
where $\mathrm{X}$ is $\mathrm{A}$ or $\mathrm{B}$, and for the
model A
\begin{equation}
F_{\mathrm{A}}(\phi_{k})\equiv\frac{(1+w)}{2}[2(\phi_{k}^{\prime\prime}-\mathcal{H}\phi_{k}^{\prime})+\square\phi_{k}-\kappa a^{2}(\rho+3p)\phi_{k}],
\end{equation}
and for the model B
\begin{equation}
F_{\mathrm{B}}(\phi_{k})\equiv\frac{(1+w)}{2}[2(\phi_{k}^{\prime\prime}-\mathcal{H}\phi_{k}^{\prime})+\square\phi_{k}].
\end{equation}
The equation (44) can be understood as a harmonic oscillator in the
expanding universe, with an external force given by $F_{\mathrm{X}}(\phi_{k})$.
First of all, we note that all the evolution converge to the results
of GR when $V\rightarrow\infty$, since the field decrease faster
to zero if the potential scale is bigger. The value of the field is
considerable only at early era of universe and soon approaches to
zero, hence we may think of the force term as an additional source
to the matter at early time. Let us investigate effects of this force
term for each model and scale. Since the conformal time can be thought
of maximal comoving photon path and $k$ denotes momentum, the behavior
for the large scale is approximated by the condition $\left|k\tau\right|\ll1$,
and for the small scale $\left|k\tau\right|\gg1$. 

First let us consider the model A. As we have discussed, $\phi_{k}$
decrease rapidly in the radiation era. So we may think of the fluid
components in the last term of (45) is only constituted by radiation
with an equation of state $w=1/3$. Hence we approximate it as $\kappa a^{2}(\rho+3p)\phi_{k}\approx2\kappa a^{2}\rho\phi_{k}=6\mathcal{H}^{2}\phi_{k}$
and find
\begin{equation}
F_{\mathrm{A}}(\phi_{k})\sim-(k^{2}+12a^{2}V_{\mathrm{A}}M_{\mathrm{P}}^{2})\phi_{k}-6\mathcal{H}(\phi_{k}^{\prime}+\mathcal{H}\phi_{k}).
\end{equation}
Notice that the former term is always negative, since in our approximation
$\phi_{k}$ is always bigger than zero. If there were no latter term,
the force term would be negative and expressed as $F_{\mathrm{A}}\propto-\phi_{k}$
and would function as a friction, which reduces the oscillation amplitude
of $\Delta_{k}$. From (43), the latter term can be expressed by $-6\mathcal{H}f^{\prime}/a$.
If this term is positive and bigger than the absolute value of the
former term, and hence $F_{\mathrm{A}}(\phi_{k})$ is positive, this
external force term can function as an additional source, which enhances
the oscillation of the photon field. This term is positive if $\int d\tau\,(k^{2}+4VM_{\mathrm{P}}^{2}a^{2}-a^{\prime\prime}/a)>0$,
hence one may think that the force term would be more likely to be
positive if we have large potential scale. However, as $V$ becomes
larger the total amount of $f^{\prime}$ decreases, so large potential
scale does not ensure the positive force term. Rather, it is positive
when we have enough small $V$ yielding noticeable amount of the force
term to affect the evolution of the field and when the above condition
is satisfied. Also, it is clear that it would be more likely to be
positive if we consider high-$k$ region or small scale rather than
small-$k$ region or large scale.

For the model B, the situation is simillar but there is important
differences. We write the force term for the model B as 
\begin{equation}
F_{\mathrm{B}}(\phi_{k})\sim-(k^{2}+12a^{2}V_{\mathrm{A}}M_{\mathrm{P}}^{2})\phi_{k}-6\mathcal{H}\phi_{k}^{\prime}.
\end{equation}
Notice again that the term appearing in the model A, $-6\mathcal{H}^{2}\phi_{k}$,
comes from the fact that the model A does not have symmetry like the
model B. It is clear that the total force term in the model B is more
likely to be positive than in the model A, when we re-write (48) as
$F_{\mathrm{B}}(\phi_{k})\sim F_{\mathrm{A}}(\phi_{k})+6\mathcal{H}^{2}\phi_{k}$.
In conclusion, we can expect that the power spectra in the model B
would be more likely to be bigger than in the model A and GR, since
we have more oscillative $k$ region of matter field contributing
to the spectra. Finally, we note once again that in the limit of $V\rightarrow\infty$
the results recovers GR consequently, by the fact that the additional
field converge to zero, which can be inferred from equation (43).
Its effects on the power spectra disappear in this limit, regardless
of any effects occuring in finite region of $V$. 
\begin{figure}
\noindent \centering{}\includegraphics[scale=0.28]{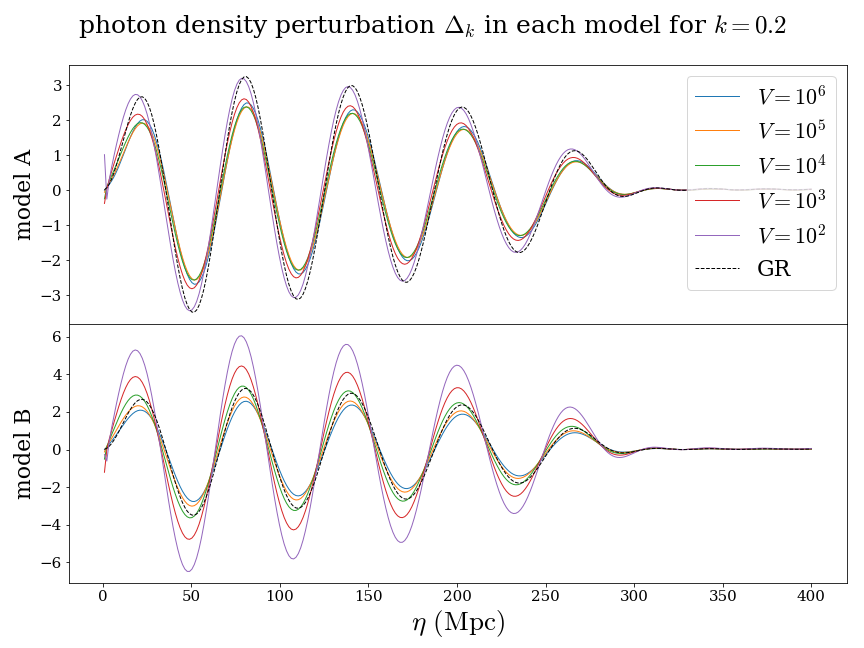}\caption{Plot for the evolution of photon energy density perturbation in each
model. Horizontal axis denotes conformal time with the unit of Megaparsec
and vertical axis denotes the value of perturbation $\Delta_{k}$.}
\end{figure}

We illustrate our discussion by plotting an evolution of the photon
energy density perturbation for an example. We plot for the case $k=0.2\,\mathrm{Mpc}^{-1}$
and various values of $V$ for each model. Note that the potential
scale is dimensionless. If one wants to understand it as field mass,
one may use the formula $m^{2}=4VM_{\mathrm{P}}^{2}$. As we expected,
we can see that from the figure 1 the amplitude of photon perturbation
depends on the potential scale. For a first few value, the amplitude
decreases as $V$ becomes smaller, but for some enough small value
it increases again. In the model B, the response to the change of
$V$ is more sensitive and there are more cases that the amplitude
is bigger than in the model A and GR. Moreover, it is worth to note
that the force term does not much change the location of peaks, or
oscillation frequency.
\begin{figure}
\centering{}\includegraphics[scale=0.28]{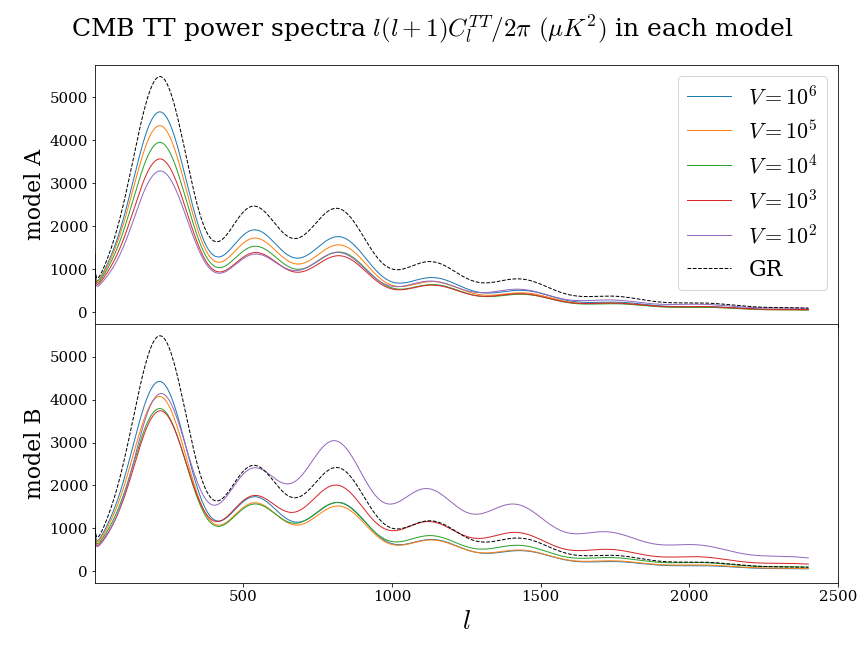}\caption{The lensed CMB TT power spectra in the unit of $\mu\mathrm{K^{2}}$
in each model.}
\end{figure}

Next we plot our main results, CMB power spectra multipoles. First
we plot TT power spectra, which is plotted in the figure 2 for each
model. The TT power spectrum is only originated from temperature anisotropy.
Let us discuss the model A first. As one can see from the upper part
of the figure 2, the additional field affects for all range of $l$.
First we mention that a few values for low-$l$ region are a little
bit bigger than the results of GR, but the difference is subtle and
it is hard to notice the difference from the graph. Rather, the effect
is dominated by high-$l$ region, which means small scale. Although
there is a small increase of the spectrum in some low-$l$ values,
in general it decreases as the potential scale become smaller unless
$V_{\mathrm{A}}<10^{3}$. This results may be inferred from the discussion
for the photon field perturbation. Since the field makes the photon
field perturbation amplitude to be changed, the CMB anisotropy intensity,
that is originated mostly from the photon perturbation, should be
affected by them. If the photon field is less oscillative for many
values of $k$, the spectra would decrease in general, and vice versa.
As we can see from the figure 1, for the model A except for $V_{\mathrm{A}}=10^{2}$
the amplitude tends to decrease and hence the spectra also decrease
overall. For the value $V_{\mathrm{A}}=10^{2}$, there is some region
which is more oscillative than in GR but this increase is minor in
the all region of $k$, and this is why we observe that the spectra
is slightly bigger than for the case $V_{\mathrm{A}}=10^{3}$ only
in the high-$l$ region. Moreover, we can see that the location of
peakes are almost not changed, and of course the results become close
to GR when $V\rightarrow\infty$.

Second, let us describe the TT spectra in the model B, which is plotted
in the below part of the figure 2. The general features are simillar
with the case for the model A, however, on the small scale there is
significant increase of the spectra. Especially, for the region $l\apprge1500$
and the value $V_{\mathrm{B}}=10^{3}$ the values become greater than
GR. This is not surprising, as we have shown that in the model B the
photon perturbation has more region of $k$, in which the amplitude
is much more bigger than in the model A or GR. In the model A, this
is not possible because the field works as a friction rather than
a source for the photon field oscillation in many region of $k$.
Whereas in the model B, we can suspect that there is an enough amount
of much oscillative high-$k$ region of the photon field that contributes
to the spectra, and that this high-$k$ region mostly contributes
to the high-$l$ region. This is the most noticeable differences between
our two models. In the limit of $V_{\mathrm{B}}\rightarrow\infty$,
however, the results recovers GR consequently, as it did in the model
A.
\begin{figure}
\centering{}\includegraphics[scale=0.28]{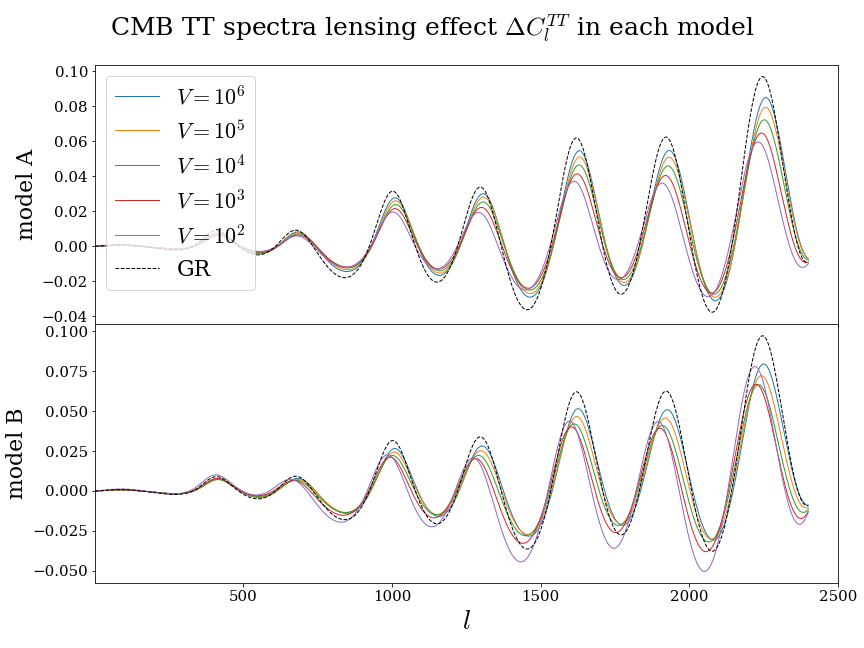}\caption{The difference between lensed spectra and raw spectra in each model.}
\end{figure}

We also plot the difference of lensed TT spectra and raw (unlensed)
TT spectra, which is given by $\Delta C_{l}^{TT}\equiv((C_{l}^{TT})_{\mathrm{lensed}}-(C_{l}^{TT})_{\mathrm{raw}})/(C_{l}^{TT})_{\mathrm{raw}}$,
in the figure 3. In the upper part of the figure, for the model A
the field weakens lensing effect and from this fact we can conclude
that the additional field makes the photons less deflected. Also,
we can see that the locations of peaks and troughs are slightly shifted
to left as the potential scale become smaller. This feature is easily
observed in the model B in the below part of the figure 3, as well.
However, there is also remarkable difference in the lensed spectra.
For high-$l$ region and low enough value of the potential scale such
as $V_{\mathrm{B}}=10^{2}$, the deviation increases, especially in
the troughs. The increases in the peaks are relatively small, and
even if the values of the troughs exceed the value of GR the values
in the peaks may not. Hence, we conclude that the contribution in
the TT spectra arising from the lensing effect becomes more important
when we consider the low potential scale.
\begin{figure}
\centering{}\includegraphics[scale=0.28]{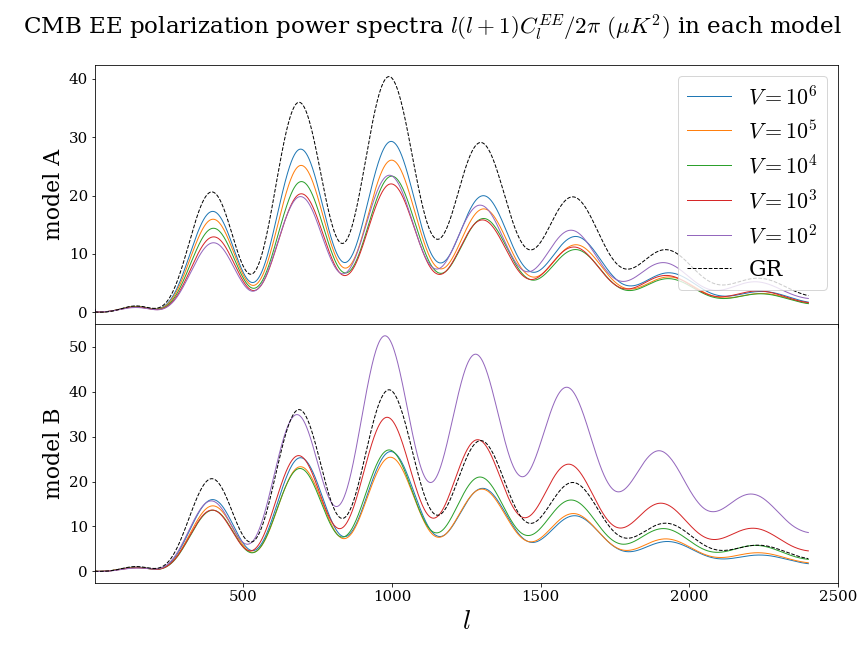}\caption{The CMB EE polarization power spectra in the unit of $\mu\mathrm{K^{2}}$
in each model.}
\end{figure}
\begin{figure}
\centering{}\includegraphics[scale=0.28]{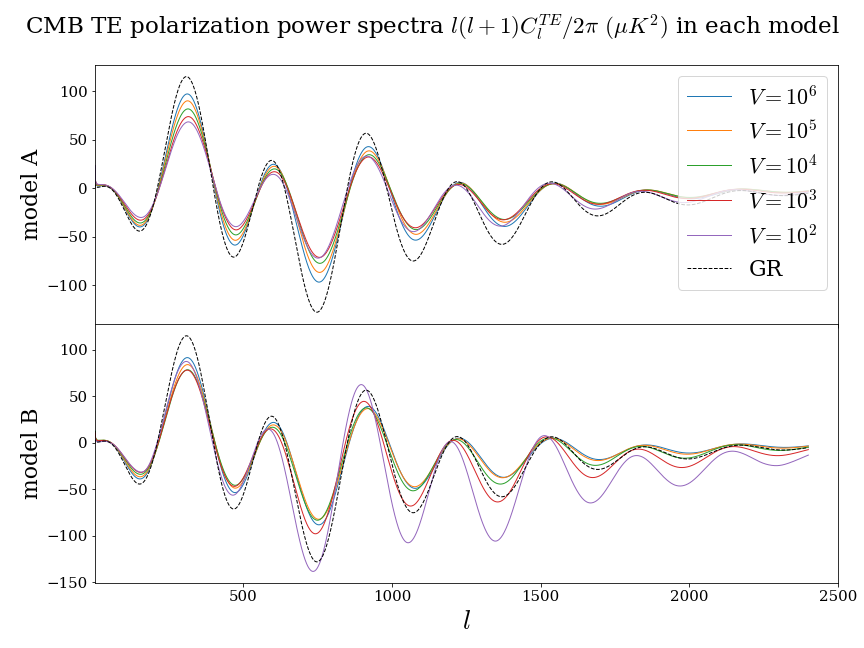}\caption{The CMB TE polarization power spectra in the unit of $\mu\mathrm{K^{2}}$
in each model.}
\end{figure}
\begin{figure}
\centering{}\includegraphics[scale=0.28]{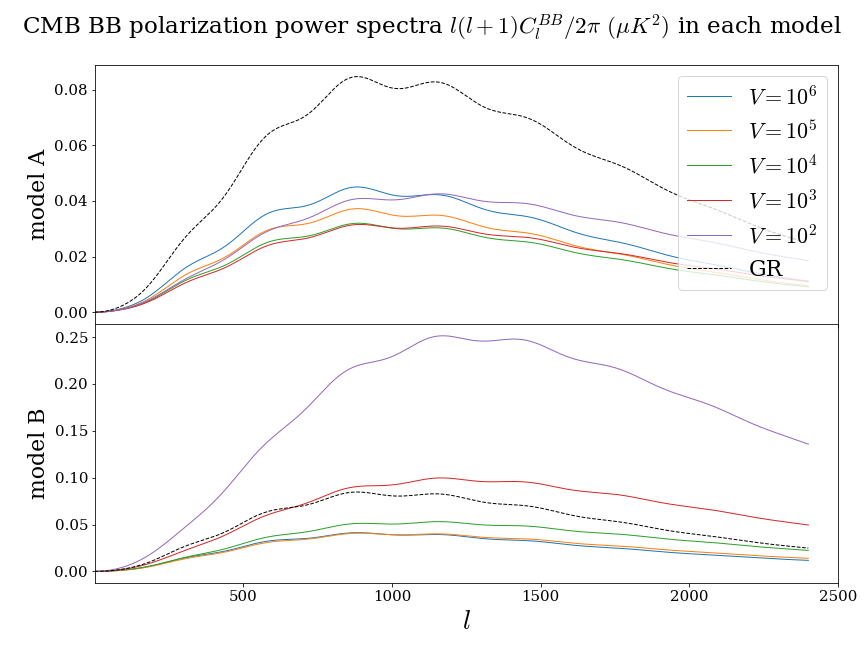}\caption{The CMB BB polarization power spectra in the unit of $\mu\mathrm{K^{2}}$
in each model.}
\end{figure}

In figure 4, 5, 6, we plot the results for EE, TE, and BB power spectra
in each model. They have almost simillar feature as TT spectra and
we only mention a few differences. In the figure 5, we can see that
the overall properties are simillar as TT spectra but the deviation
is much more concentrated in the troughs, especially in the high-$l$
region. Another feature we also have to concern is that the deviation
of BB spectra seems to be much more bigger than the other spectra.
This phenomena is more drastic in the model B, but we can see that
the smaller potential scale can give larger spectra also in the model
A, when comparing $V_{\mathrm{A}}=10^{2}$ and $V_{\mathrm{A}}=10^{3}$.
Let us describe the feature more in detail. As the value of the potential
scale decreases, BB spectra first decreases in general. For some enough
value of the potential scale, however, the spectra increase overall
and for the model B values exceeding GR are also obverved. Note that
this feature of BB spectra comes from the lensing. Of course, BB spectra
can be also made from tensor perturbation and one may think that there
would be some differences in evolution of gravitational waves and
this changes would affect BB spectra. In our models, however, there
is no differences from GR in tensor perturbation itself. This effect
solely comes from the change of lensed B modes. The amount of deviation
in BB spectra from lensing makes no big difference from the deviation
in TT spectra lensing, but the original BB spectra from GR has much
smaller value than TT spectra and hence the proportion of deviation
is much bigger in BB spectra. Therefore, this discussion on lensing
effect suggest that we have to pay attention on lensing, especially
when we study B modes.

Let us make some comparison between our two models and the other researches.
There are many exhaustive studies on CMB theory and observation for
modified gravity theories, and comparisons for all these models are
outside the scope of this paper. For a brief information and numerical
analysis of some representative modification in CMB theory, see {[}23{]}.
Here we will only compare some of them briefly. First, we mention
ekpyrotic scenario {[}24{]}, which claims that the big bang is triggered
by big bounce. This may be regarded to be somewhat simillar with our
models, in the sense that it also deals with the very early era of
the universe. However, whereas ekpyrotic scenario does not produce
primordial gravitational waves, our two models both do not affect
the existence of gravitational waves. It is purely determined by the
inflationary universe models.

Second, let us explain differences between our models and many modification
of EM tensor in \textgreek{L}CDM model. These models, such as quintessence,
are often proposed to resolve cosmic constant problem. Many times
they result in different properties of dark energy in the universe,
and typically affect Integrated Sachs-Wolfe effect, which causes the
low-$l$ region of spectra to be changed. One may also think of our
models as modifications of matter, since we have extra EM tensor and
there is actually no change in gravity itself. However, in our models
new EM tensor only appears in perturbative scale and has no effects
in the background. Rather, the perturbation of the new scalar field
has strong impact on the oscillation of matter perturbation at the
early times of the universe.

Third, we discuss some differences between CMB in BD theory and our
models. A pioneering work for CMB in BD theory which applies covariant
formalism is given by {[}25{]}. Our models are of course originated
from BD theory. However, due to the symmetry breaking, non-mininal
coupling of gravity and the scalar field vanishes and this brings
a lot of differences. For example, in BD theory the deviation of CMB
TT spectra intensity from GR is quite negligible, comparing with the
deviation of the location of peaks. In our models, however, there
is almost no deviation of the location of peaks. Since the scalar
field alters the amplitude of photon field perturbation, it also changes
the intensity of CMB power spectra. We also emphasize especially that
the model B has another extraordinary feature, that is an increase
of the intensity at the very small scale ($l\gtrsim1500$), which
makes it larger than the result of GR for the small enough potential
scale. This is quite distinguishable difference from other researches.

Finally, we finish this section by discussing a few methods for examining
our results with observational data. As we have discussed earlier,
our models only can be verified at perturbative level and we already
know many tests which apply perturbation theory. But we also have
to consider the fact that the scalar field perturbation rapidly decrease
in the radiation era. This implies that we cannot verify our results
only by considering perturbations at the current era. Hence, it is
necessary to also consider the early era of the universe and of course
CMB observation provides a best way for this. Therefore, here we have
to only discuss tests satisfying this condition. First of all, data
from Planck satelite are most representative observational results
for analyzing CMB power spectra. But there are some other observational
tests that would be useful when comparing with observation. For example,
Lyman-$\alpha$ forest observations shows many absorbtion lines in
quasar spectra, and it may be helpful for analyzing high-$k$ density
fluctuations, in which our models differ from GR {[}26{]}. Many recent
observations such as DES (Dark Energy Survey) {[}27{]}, SDSS (Sloan
Digital Sky Survey) {[}28{]} would be also helpful, showing how fluctuations
of dark matter and dark energy have evolved. Also, currently there
are many experiments on CMB which focus on small angular scale of
CMB spectra. For example, AdvACTPol (Advanced Acatama Telescope) {[}29{]},
SPT-3G (South Pole Telescope) {[}30{]}, Simons Observatory {[}31{]},
and CMB-S4 {[}32{]} would be useful to examine the validity of our
models.

\section{Conclusion}

In this paper, we proposed a way for probing two kinds of primordial
symmetry breaking with CMB power spectra. The first model (the model
A) is originated from Zee's broken-symmetric theory of gravity, and
the second model (the model B) comes from applying Palatini formalism
to the model A. Especially, symmetry breaking phenomena in the model
B is appeared to have geometrical feature. We presented new EM tensors
(28) and (33) for the perturbation theories in each model, to show
a way to verify the models with observation.

To plot the computed results we used the numerical code CAMB. However,
for the numerical reason to speed up the code we had to apply specific
approximation scheme for the large potential scale. The scalar field
in new EM tensors appeared to affect the evolution of ordinary matter
perturbation, and it decrease or increase the amplitude of the perturbation.
This also brings differences in the intensity of CMB power spectra.
For the model B, we discovered that the increase in the high-$l$
region of the spectra is more drastic than in the model A. We also
compared our results with the other studies and mentioned some observational
tests that would be helpful when verifying our models.

Our studies have some noticeable meanings. The models we proposed
open a new way for probing primordial symmetry breaking through cosmic
observation. Our models do not concern mechanisms like branching of
gravity from other forces, the current scenario explaining appearances
of four forces. But at least it might give some clue for the other
way of probing symmetry breaking phenomena in the universe, using
cosmological method.

However, our results are restrictive and require more researches.
First, we restricted ourselves to some simple cases. We only considered
simplest scalar-tensor theory and did not consider torsion. Weyl geometry,
which motivated our studies, is actually a simple case without torsion
of Lyra's geometry. Although it is hard to apply covariant formalism
to theories with torsion since one may cannot choose proper foliation
in this case and hence cannot define 4-velocity and kinematical quantities,
many high-energy gravity theories in these days concerns torsion so
we need to develop appropriate mathematical formulations for them.

Second, we think that our study only should be considered as some
kind of examples illustrating one possibility for cosmological observation
of primordial symmetry breaking. Our models presume a certain type
of physics before the inflation and at the Planck scale, on which
we cannot speak precisely unless we establish valid quantum gravity
theory. No one knows what kind of symmetry would govern the spacetime
at the Planck scale and how big the mass of a field invoking symmetry
breaking would be. Nevertheless, we believe that our study provides
some conceptual usfulness. Even though the problem mentioned above
are not resolved, the limit of the value $m$ (or $V$) beyond the
Planck mass can be understood as the upper limit, below which a scalar-tensor
theory becomes GR as its effective theory because the influence of
the additional EM tensor diminishes.

Finally, our numerical approximation is almost safe but not accutate.
We have to find another way to overcome the numerical tackles. Of
course this is inevitable when it comes to comparing our theory with
observational data such as Planck setellite data. Moreover, it would
be some helps for more detailed studies on differences between two
models or the other models based on different type of symmetry. We
also think that we need more concentrative studies for the density
perurbation of various matters, the lensing effect and its contribution
to the spectra, to make our theory contact with the observation. Comprehensive
researches, including comparison with the observational data, might
show us more interesting pictures of the theory. The careful study
with the observational tests mentioned above, are waiting to be performed.
\begin{acknowledgments}
We thank to anonymous referee for many important comments, especially
for correcting our uncertain description of the first manuscript on
BB spectra and gravitational waves, and helping the overall improvement
of the manuscript. We also thank to Jonghyun Sim for helpful comments.
This research was supported by the Basic Science Research Program
through the National Research Foundation of Korea (NRF) funded by
Ministry of Education, Science and Technology (NRF-2017R1D1A1B06032249).
\end{acknowledgments}


\begin{thebibliography}{10}
\bibitem{key-21}The code slows approximately $\sim10^{4}$ times
or more when the deviation of the multipole $C_{l}$ is less than
the order of $\sim10^{-3}$ under the standard 8-core personal computer
when $V\apprge10^{3}$.
\end{thebibliography}
\end{document}